\documentclass[preprint,
 amsmath,amssymb,
 aps,
showpacs,floats,
longbibliography
]{revtex4-1}
\usepackage[table]{xcolor}
\usepackage{graphicx,color}% Include figure files
\usepackage{array}
\usepackage{multirow}
\newcommand{\head}[1]{\textnormal{\textbf{#1}}}

\begin{document}

%\preprint{APS/123-QED}

%\title{Rheology of immersed and dry frictional spheres}
\title{The influence of surface roughness on the rheology of immersed and dry frictional spheres}

\author{Franco Tapia, Olivier Pouliquen, and \'Elisabeth~Guazzelli$^{\ast}$}

\affiliation{Aix-Marseille Univ, CNRS, IUSTI, Marseille, France\\$^{\ast}$ now at Laboratoire Mati\`ere et Syst\`emes Complexes, CNRS UMR 7057, Universit\'e de Paris, 75205 Paris Cedex 13, France}

\date{\today}% It is always \today, today,

\begin{abstract}

Pressure-imposed rheometry is used to examine the influence of surface roughness on the rheology of immersed and dry frictional spheres in the dense regime. The quasi-static value of the effective friction coefficient is not significantly affected by particle roughness while the critical volume fraction at jamming decreases with increasing roughness. These values are found to be similar in immersed and dry conditions.  Rescaling the volume fraction by the maximum volume fraction leads to collapses of rheological data on master curves. The asymptotic behaviors are examined close to the jamming transition.
\end{abstract}

%\pacs{}

\maketitle

\section{Introduction}
\label{sec:intro}

While being quite different particulate systems, viscous non-colloidal suspensions and dry granular materials can be described by rheological laws which use a common framework \cite{BoyeretalPRL2011}. When a dense collection of dry hard spheres (having diameter $d$ and density $\rho_p$) is sheared at a given shear rate, $\dot{\gamma}$, under an imposed particle pressure, $P$, the rheology is determined by the knowledge of two dimensionless quantities: the packing faction, $\phi$, and the effective friction coefficient (or stress ratio), $\mu=\tau/P$, where $\tau$ is the shear stress. Dimensional analysis implies that these two quantities only depend on a single inertial dimensionless number, $I= d \dot{\gamma} \sqrt{\rho_p/P}$, and are thus written as $\mu(I)$ and $\phi(I)$, see e.g. \cite{ForterrePouliquen2008}. A similar approach can be applied to the viscous flow of suspensions of hard non-Brownian spheres. The rheological laws adopt a similar form, $\mu(J)$ and $\phi(J)$, but with the inertial number, $I$, replaced by a viscous number, $J=\eta_f \dot{\gamma}/P$, where $\eta_f$ is the suspending fluid viscosity \cite{BoyeretalPRL2011}. This frictional formulation is equivalent to the more classical description of the rheology of suspensions in terms of effective viscosities: the shear, $\eta_s(\phi)=\tau/\eta_f \dot{\gamma}=\mu/J(\phi)$, and normal, $\eta_n(\phi)=P/\eta_f \dot{\gamma}=1/J(\phi)$, relative viscosities where $J(\phi)$ is the inverse function of $\phi(J)$ which is perfectly defined since $\phi(J)$ is monotonic.

These rheological properties become singular in the dense regime when reaching the jamming transition for which the particulate system ceases to flow, both in the viscous (suspensions) and inertial (dry granular materials) cases. There is not yet a complete understanding of these singular behaviors and the current description remains rather empirical. The major problem lies in relating the mechanics at the grain scale to these macroscopic properties. As the jamming transition is approached, particles form an extended network of contacts, see e.g. \cite{Catesetal1998,Lerneretal2012,Andreottietal2012}, and the rheology is then dominated by contact forces, even in the case of viscous suspensions for which the hydrodynamics interactions between the particles become of lesser importance and are overshadowed by direct contact interactions, see e.g. \cite{GuazzelliPouliquen2018}. Numerical simulations and scaling arguments \cite{DeGiulietal2016,Trulssonetal2017} have recently pointed out the role of friction between the grains, and in particular the effect of varying the interparticle sliding friction coefficient, $\mu_{sf}$. For low $\mu_{sf}$, the dissipation mainly occurs in the interstitial fluid between the grains for viscous flow of hard spheres while it is due to inelastic particle collisions for inertial flow. Similar physical mechanisms occur at large $\mu_{sf}$ in a rolling regime wherein the spheres roll relative to each other. The dissipation due to the sliding contacts is dominant in the intermediate range of $\mu_{sf}$ which has been expected to be most relevant experimentally in particular close to the jamming transition.

Only a few experiments have examined the impact of interparticle friction, and in particular surface roughness, on the rheological properties of these particulate systems. The earliest study of Lootens \textit{et al.} \cite{Lootensetal2005} showed that increasing surface roughness of colloidal particles shifts the shear-thickening transition to lower critical stresses. For non-colloidal viscous suspensions, the available experimental results show that increasing roughness leads to higher viscosities \cite{Moonetal2015,TannerDai2016} in qualitative agreement with numerical simulations \cite{Gallieretal2014,Marietal2014} using rather large value of $\mu_{sf}$ ($\approx0.5-1$) to match the observations. The influence of the interparticle friction has been mostly studied in numerical simulations for dry granular media \cite{SunSundaresan2011,Chialvoetal2012,DeGiulietal2016}. No systematic experimental study has been undertaken in the very dense regime close to the jamming transition. The present work aims at filling this gap by using pressure-imposed rheometry which is particularly suitable to infer the singular rheological properties of particulate systems at the jamming transition \cite{BoyeretalPRL2011,Kuwanoetal2013,Dagois-Bohyetal2015,Falletal2015,Tapiaetal2017}. These rheological measurements are reported in section~\ref{sec:results}, both for dry and immersed hard spheres, after presenting the materials and techniques in section~\ref{sec:exp}; conclusions are drawn in section~\ref{sec:conclusion}.

\section{Experiments}
\label{sec:exp}

\subsection{Rheometry}

\begin{figure}
\begin{center}
\includegraphics[width=0.8\textwidth]{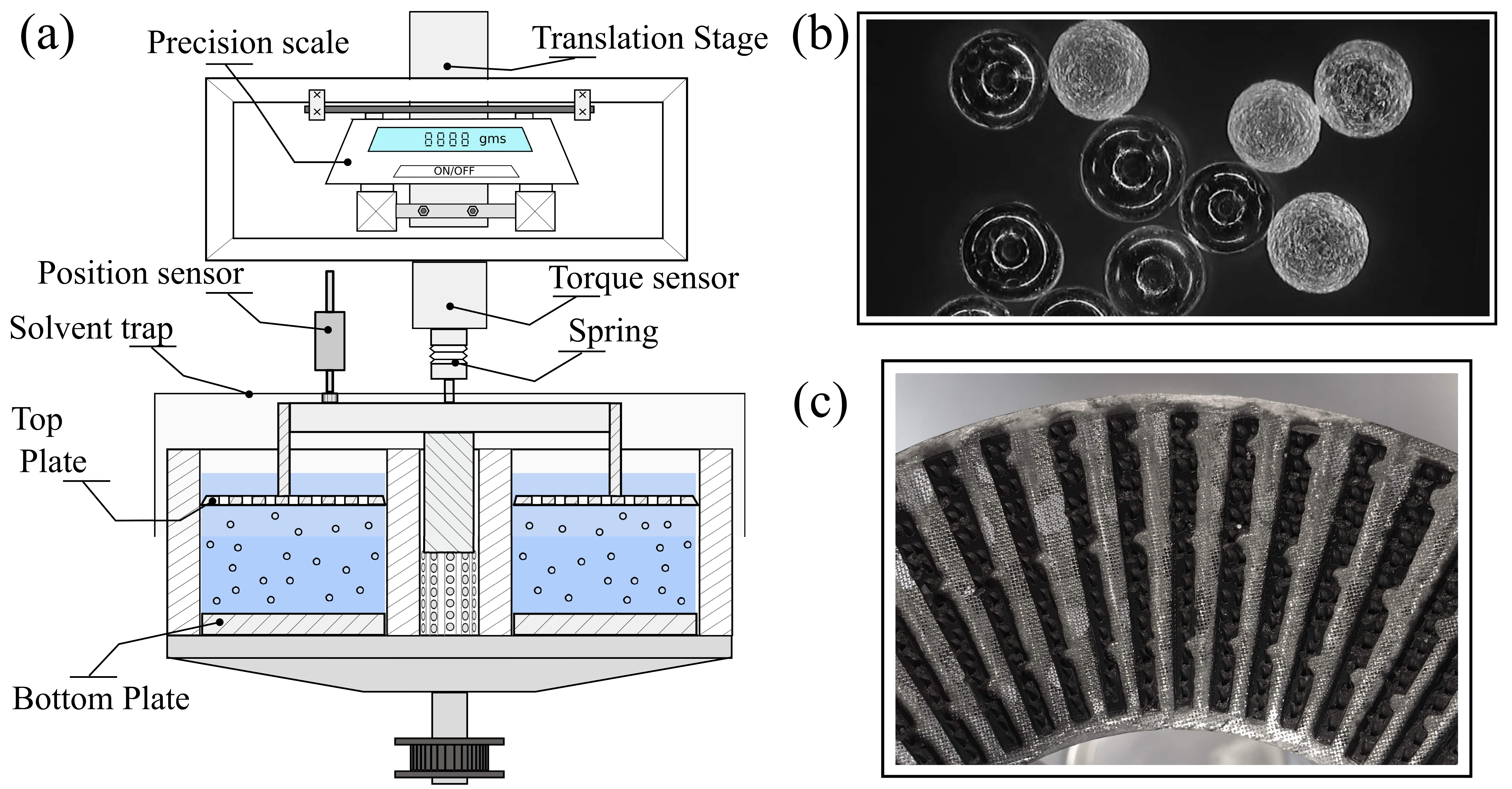}
\caption{(a)~Sketch of the experimental apparatus. (b)~Microscopic image of slightly and highly roughened polystyrene spheres.  (c)~Image of the top permeable plate showing the corrugated strips.}
\label{fig:Rheo} 
\end{center}
\end{figure}

The experimental apparatus, depicted in Fig.\,\ref{fig:Rheo}(a), is a custom-made rheometer enabling pressure- and volume-imposed rheological measurements of dry or immersed granular materials. The granular sample is subjected to a simple shear in a plane-plane geometry consisting of a cylindrical annulus (of internal and external radii $R_1 = 43.95$\,mm and $R_2 = 90.28$\,mm, respectively) that is attached to a bottom plate and is covered by a top plate. In order to obtain a linear shearing of the granular sample, the cylindrical annulus rotates at a constant angular velocity controlled by an asynchronous motor (Parvalux SD18) regulated by a frequency controller (OMRON MX2 0.4 kW) while the top plate does not rotate. A wide range of shear rate, $\dot{\gamma}$, can be achieved, spanning between $0.02$ and $130$\,s$^{-1}$. The important feature of this rheometer is that the top plate is permeable to the fluid but not to the particles, see a blowup of a picture of this plate in Fig.\,\ref{fig:Rheo}(c). It is manufactured with holes of size $2-5$\,mm enabling fluid to flow through it but is also covered by a $0.24$\,mm nylon mesh designed to stop particles. This top plate can be moved vertically by using a linear positioning stage (Physics Instrumente M-521) and fits into the cylindrical annulus with a precision of $280$\,$\mu$m. Both the top and bottom plates are also roughened by positioning corrugated strips of height and width 0.5\,mm onto their surfaces, as seen in Fig.\,\ref{fig:Rheo}(c). This apparatus was initially built to study suspension rheology \cite{Dagois-Bohyetal2015,Tapiaetal2017} and has been adapted to make possible the investigation of dry granular material. An important ingredient was to create appropriate roughnesses on the top and bottom plates to enable bulk granular motion. Another important point was to operate the rheometer in an air-conditioned room (at $25^{\circ}$ Celcius)  with a high level of humidity (at a relative humidity of 80\%) to avoid electrostatic effects between the dry polystyrene particles.

The shear stress, $\tau$, is deduced from the torque exerted on the top plate measured by a torque transducer (TEI - CFF401) connected to the top plate. The component of the normal stress perpendicular to the top plate, simply referred as the particle pressure $P$, is given by a precision scale (Mettler-Toledo XS6002S) attached to the translation stage. The bulk packing fraction of the sample, $\phi$, can be adjusted by displacing the top plate. The plate position, $h$, is continuously measured by a position sensor (Novotechnik T-50). A feedback control system connects the positioning stage and the precision scale in order to perform pressure-imposed experiments on the sample. In this pressure-imposed mode, the resulting shear stress, $\tau$, and packing fraction, $\phi$, are measured as functions of shear rate, $\dot{\gamma}$, for a set particle pressure, $P$, once steady state is established. Classical volume-imposed rheometry can also be performed by fixing the top plate position, i.e. fixing the volume fraction $\phi$. In this volume-imposed mode, the shear stress, $\tau$, and particle pressure, $P$, are measured as functions of shear rate, $\dot{\gamma}$, for a given volume fraction $\phi$. Note that a soft spring is positioned between the top plate and the torque sensor to avoid blockage during highly dense experiments. A series of calibration experiments with a pure fluid is also performed to infer the undesired friction from the central axis as well as the viscous contribution from the gap between top plate and cell walls.  These effects are subtracted to the torque measurements. Buoyancy effects are also accounted for in the measurements of the normal force that the particles exert on the porous plate in the gradient direction.
 
In immersed experiments performed in pressure- or volume-imposed modes, polystyrene particles, shown in Fig.\,\ref{fig:Rheo}(b), are suspended in a Newtonian fluid (polyethylene glycol-ran-propylene glycol monobutylether) which has a matching density with the particles ($\rho_f = 1056$\,kg/m$^3$ at $25^{\circ}$ Celcius) and a large viscosity ($\eta_f=2.01$\,Pa\,s at $25^{\circ}$ Celcius). Air bubbles are removed by vacuum extraction and then by a long period of pre-shearing. In experiments with dry particles undertaken in the pressure-imposed mode, the cell is filled up to a given height by a rain-like procedure which provides a good homogeneous distribution of the particles within the cell. A small dispersion in size is required to avoid ordering. Ranges of  imposed pressure ($P = 60-380$\,Pa) and sample height ($h<16$\,mm) are chosen to avoid shear banding nucleation. It is important to stress that the smallest confining pressure has to be larger than the hydrostatic pressure to avoid large inhomogeneity in the bulk. This results in a more limited range of $\phi$ in the dry case.

\subsection{Manufacturing rough spheres}

\begin{figure}
\begin{center}
\includegraphics[width=0.8\textwidth]{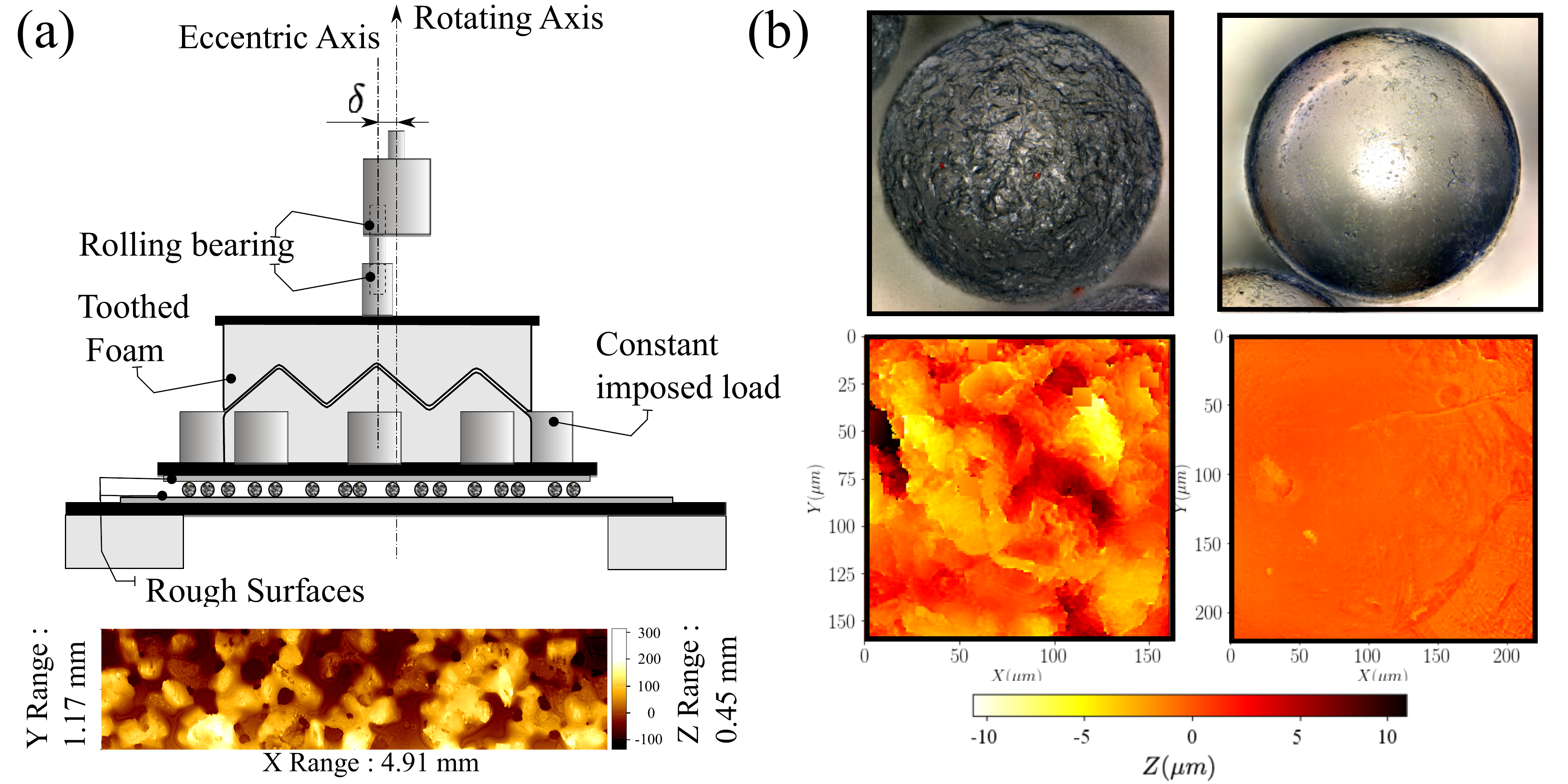}
\caption{(a)~Top: Sketch of the particle roughening apparatus. Bottom: Confocal image of the upper rough surface.   (b)~Microscopic images of typical slightly (right) and highly (left) roughened spheres. Bottom: Corresponding confocal images of the particles surface.}
\label{fig:roughening} 
\end{center}
\end{figure}

Two set of spheres having different surface roughnesses but approximatively the same size are used in the experiments, see Figs.\,\ref{fig:Rheo}(b) and \,\ref{fig:roughening}(b). The particles labeled ``Slighly Roughened'' (SR) correspond to rigid polystyrene particles (Dynoseeds TS 500) manufactured by Microbeads SA, having a density $\rho_p=1050$\,kg/m$^3$ and an approximately Gaussian distribution in size with a mean diameter of $d=580$\,$\mu$m. As can be seen in the right panel of Fig.\,\ref{fig:roughening}(b), these spheres are not perfectly smooth but possess some small surface roughnesses. The particles labeled ``Highly Roughened''  (HR) are produced by further roughening these polystyrene spheres, see the left panel of Fig.\,\ref{fig:roughening}(b).

The roughening procedure consists in forcing a continuous motion of the particles between two parallel rough plates resulting in a mechanical erosion of the particle surfaces. The particle-roughening apparatus is sketched in Fig.\,\ref{fig:roughening}(a). Sandpapers (Walfcraft 80) cover both the top and bottom circular plates ($20$ cm in diameter) to avoid slippage. The bottom circular plate is fixed while the displacement of its top counterpart is driven by an electrical stirring device having a shifted rotational-axis ($\delta=5$\,mm) with respect to the bottom-plate axis. Two rolling bearings transfer rotation into translation resulting in a circular movement. Circular translation instead of rotation ensures that the particle trajectory is independent of its location. Each particle thus experiences the same mechanical erosion. In addition, the top plate exerts a static load (1200\,g) in order to amplify the impact on the particle surface. A toothed soft foam is also used to transfer the motion of the stirring device to the top plate while avoiding particle fracturing.

A typical protocol consists in spreading $1$\,g of polystyrene spheres onto the bottom plate. Fixing the stirring frequency at $1.2$\,Hz results in a relative velocity between the plate of $37$\,mm/s. Fig.\,\ref{fig:roughening}(b) shows the effect of one hour of erosion for a representative individual sphere. A slightly reduction in particle size is observed resulting in a slightly smaller mean diameter for the HR spheres, $d=540$\,$\mu$m.

\subsection{Characterizing rough spheres}

\begin{table}
\begin{ruledtabular}
\begin{tabular}{cccccccc}
&  $R_a$ ($\mu$m) &$R_q$ ($\mu$m) & $R_z$ ($\mu$m) & $\mu^d_{sf}$ &$\mu^i_{sf}$ & $\mu^d_{rf}$ & $d$ ($\mu$m)\\
\hline
SR  & $0.287\pm 0.008$  & $0.387\pm 0.008$ & $2.073\pm 0.008$              &  $0.23\pm 0.03$  &  $0.25\pm 0.03$ &   $0.004\pm 0.001$ &       $580\pm20$ \\
HR  &  $1.896\pm 0.008$ & $2.410\pm 0.008$  &             $9.808\pm 0.008$               & $0.37\pm 0.03$   & $0.35\pm 0.03$  & $0.007\pm 0.001$ &      $540\pm20$\\
\end{tabular}
\caption{Properties of the ``Slightly Roughened'' (SR) and ``Highly Roughened'' (HR) spheres: Roughness coefficients ($R_a, R_q, R_z$), sliding friction coefficients ($\mu^d_{sf}$ and $\mu^i_{sf}$ in the dry and immersed cases, respectively), rolling friction coefficients ($\mu^d_{rf}$ in the dry case), and diameters ($d$).}
\label{tab:surface}
\end{ruledtabular}
\end{table}

To provide some indications regarding the particle geometry, typical surface-roughness characteristics have been measured, see Fig.\,\ref{fig:roughening}(b). Confocal scanning microscopy of specimens of SR and HR spheres has been performed on a surface region of $170\times170\,\mu$m$^2$. Average roughness, $R_a$, standard deviation, $R_q$, and ten-point mean roughness, $R_z = (\mid Z_{+}\mid + \mid Z_{-}\mid)$ (where $Z_{+}$ and $Z_{-}$ are the average of the 5 tallest peaks and the 5 lowest valleys, respectively), are computed after fitting and then subtracting the spherical shape. Surface properties of both batches of spherical particles are summarized in Table\,\ref{tab:surface}.
 
\begin{figure}
\begin{center}
\includegraphics[width=0.7\textwidth]{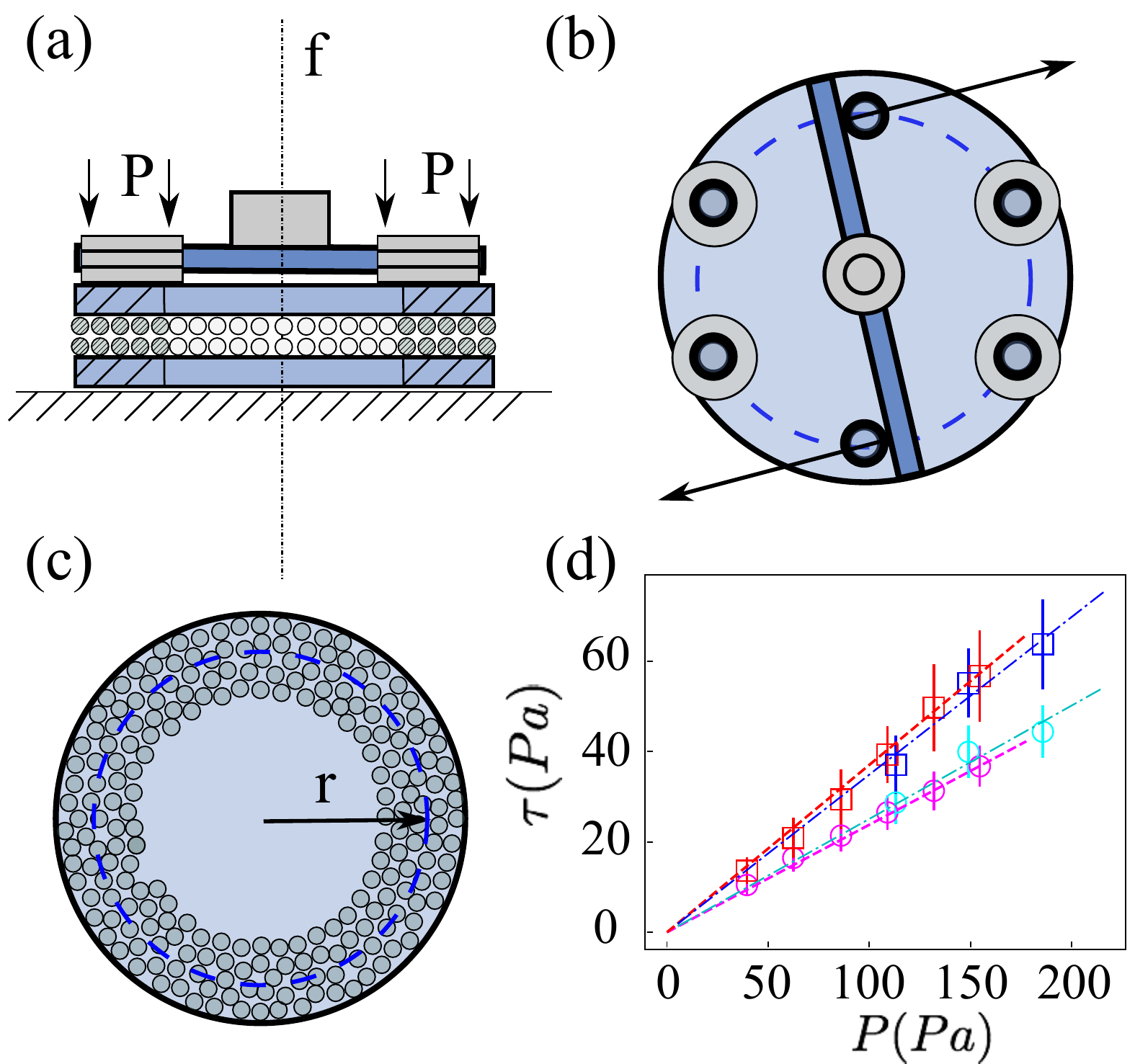}
\caption{Interparticle sliding friction experiment. (a)~ Lateral view of the setup  (b)~Top view showing the static-weight distribution on the top plate.  (c)~Bottom plate: the sample is a ring of effective radius $r=(r_2+r_1)/2 =18.5$\,mm ($r_1=25$\,mm and $r_2=12$\,mm). (d)~ Plot of the shear stress $\tau$ versus the normal pressure $P$ and corresponding linear fits for SR ($\bigcirc$) and HR ($\square$) spheres in the dry (magenta and red, respectively) and immersed (cyan and blue, respectively) cases.}
\label{fig:slidingfriction} 
\end{center}
\end{figure}
 
To characterize the interparticle sliding friction coefficients, we perform controlled sliding-experiments using two parallels plates coated with a monolayer of polystyrene particles identical to those used in the rheological experiments. The beads are glued on both the top and bottom plates in a circular annulus geometry, as shown in Fig.\,\ref{fig:slidingfriction}(a),(c).  Distributing spatially the glued particles in a random manner ensures to avoid any geometrical match between the facing surfaces during sliding. Frictional characteristics are explored by changing the sliding velocity ($8 \leqslant v \leqslant160$ mm/s) and the static load ($40 \leqslant P \leqslant 180$ Pa) imposed onto the sample by adding weights on the top plate. The resulting shear stress, $\tau$, is obtained by measuring the torque using a rheometer measuring head (Anton Paar DSR 502). This head is coupled to an horizontal rail which transfers the angular velocity to the top plate, see Fig.\,\ref{fig:slidingfriction}(a)-(b). The bottom plate is fixed while top plate rotates. Controlled experiments with a misaligned axis have been performed in order to verify that the contact area does not have any influence on the torque measurements. We also check that there is no shear-rate dependence in $\tau$. The measured shear stress, $\tau$, is plotted as a function of the applied normal stress, $P$, in Fig.\,\ref{fig:slidingfriction}(d) for two batches of SR and HR spheres. Clearly, a linear dependence is observed. A linear fit yields the slope, i.e. the sliding friction coefficient, for each sample. The interparticle sliding friction coefficients for SR and HR spheres are given in Table\,\ref{tab:surface} in both dry and immersed situations, $\mu^d_{sf}$ and $\mu^i_{sf}$ respectively. The sliding friction coefficient increases with increasing roughness but does not change significantly between dry or immersed conditions. 
\begin{figure}
\begin{center}
\includegraphics[width=0.9\textwidth]{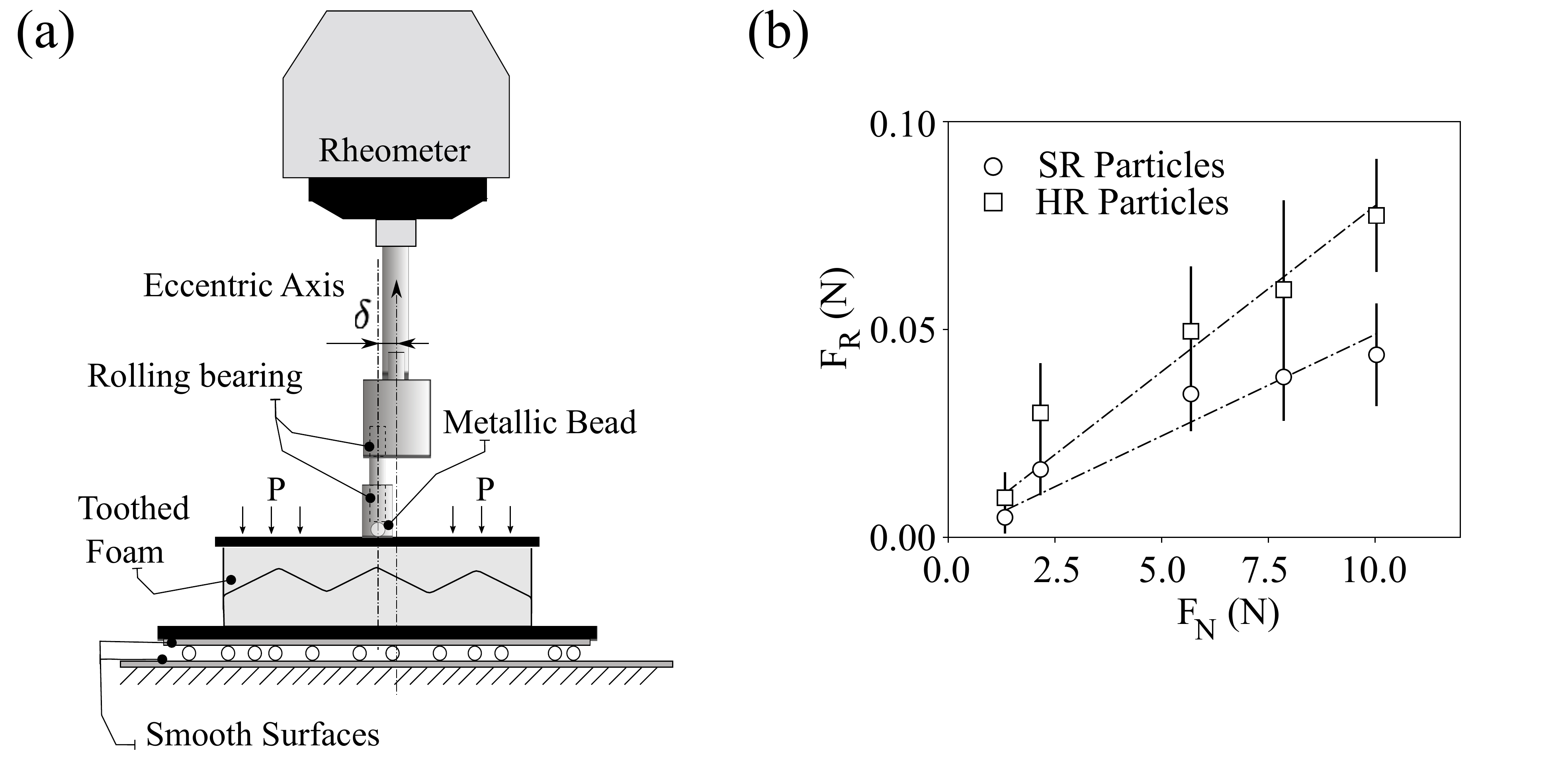}
\caption{Interparticle rolling friction experiment. (a)~ Lateral view of the setup  (b)~ Plot of the rolling force, $F_R$, versus the applied normal force, $F_N$, and corresponding linear fits for SR ($\bigcirc$) and HR ($\square$) spheres in a dry situation.}
\label{fig:rollingfriction} 
\end{center}
\end{figure}

In order to estimate the interparticle rolling friction coefficients, we also conducted controlled rolling-experiments using a device with a design similar to the particle-roughening apparatus of Fig.\,\ref{fig:roughening}(a). The bottom plate is fixed while the top plate performs a translational motion with a shifted rotational-axis ($\delta=5$\,mm) with respect to the bottom-plate axis, see Fig.\,\ref{fig:rollingfriction}(a). The particles which are sandwiched between these two plates experience the same rolling motion. The surfaces of the plates have been chosen to be rather smooth to avoid any impact on the measurements. The rolling force, $F_R$, exerted on the particles is obtained by measuring the torque using a rheometer measuring head (Anton Paar DSR 502).  This force is plotted as a function of the applied normal force, $F_N$, in Fig.\,\ref{fig:rollingfriction}(b) for the SR and HR spheres. The interparticle rolling friction coefficient, $\mu^d_{rf}$, increases with roughness but is found to be always much smaller than the sliding friction coefficient, see Table\,\ref{tab:surface}.

\section{Experimental Results}
\label{sec:results}

\begin{figure}
\begin{center}
\includegraphics[width=0.95\textwidth]{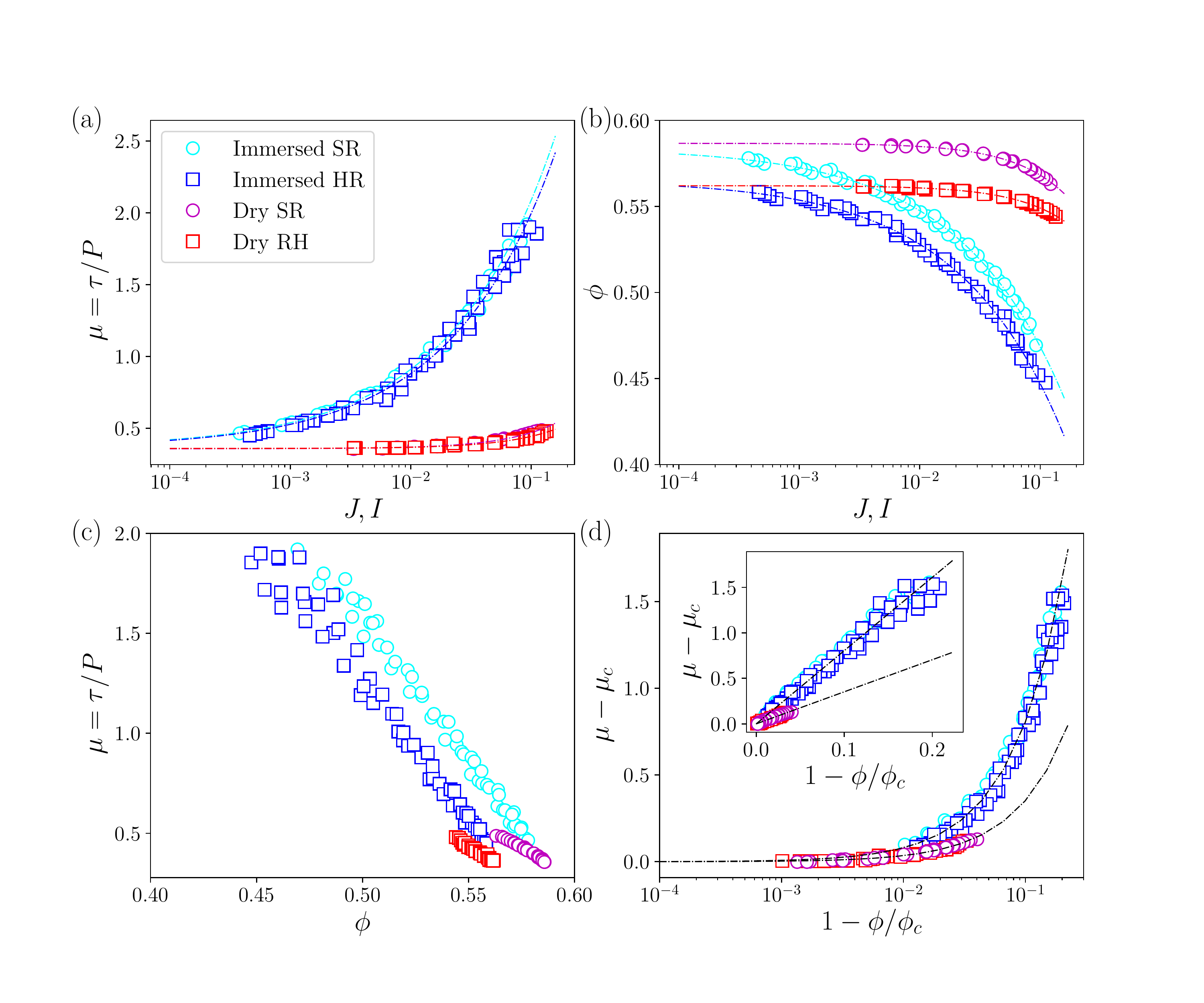}
\caption{(a)~Effective friction coefficient, $\mu=\tau/P$, and (b)~volume fraction, $\phi$, versus the viscous number, $J=\eta_f \dot{\gamma}/P$, for the immersed case (blue and cyan colors)  and versus the inertial number, $I= d \dot{\gamma} \sqrt{P/\rho_p}$, for the dry case (red and magenta colors), for SR ($\bigcirc$) and HR ($\square$) spheres. The dashed lines are the linear regressions in $\sqrt{J}$ (immersed case) and in $I$ (dry case), see Table\,\ref{tab:coeff}. (c)~Effective friction coefficient, $\mu=\tau/P$, versus volume fraction, $\phi$. (d)~Rescaled friction coefficient, $\mu-\mu_c$, versus rescaled volume fraction, $1-\phi/\phi_c$. The inset shows the same data in linear scales. The black dashed lines are the linear regressions in $1-\phi/\phi_c$ for the dry and the immersed cases.}
\label{fig:MuPhi} 
\end{center}
\end{figure}

\begin{table}
\begin{ruledtabular}
\begin{tabular}{c c c c c}   
  &          models &    label   &     $\mu_c$     &       $\phi_c$\\
\hline
\multirow{ 2}{*}{\head{Viscous (immersed)}}   &  $\mu- \mu_c \propto J^{1/2}$  &  SR      & $0.37\pm 0.01$        & $0.584 \pm 0.002$ \\
                                    &   $\phi_c-\phi \propto J^{1/2}$     & HR      & $0.36\pm 0.01$       & $0.565 \pm 0.002$  \\
\hline                                
\multirow{ 2}{*}{\head{Inertial (dry)}}  &     $\mu - \mu_c \propto I$           & SR      & $0.36\pm 0.01$          &$0.587 \pm 0.002$ \\  
                                                           &        $\phi_c-\phi \propto I$         & HR         & $0.36\pm 0.01$          & $0.563 \pm 0.002$ \\ 
\end{tabular}
\caption{Values of $\mu_c$ and $\phi_c$ for the SR and HR spheres in the immersed and dry cases.}
\label{tab:coeff}
\end{ruledtabular}
\end{table}

We display first  the rheological measurements in Fig.\,\ref{fig:MuPhi}(a) and (b) by plotting the effective friction coefficient, $\mu=\tau/P$, and the bulk volume fraction, $\phi$, as a function of the viscous number, $J=\eta_f \dot{\gamma}/P$, for the immersed particles and as function of the inertial number, $I= d \dot{\gamma} \sqrt{\rho_p/P}$, for the dry particles. A remarkable result is that $\mu$ is not significantly affected by an increase in particle roughness (or in interparticle friction). Conversely, there is a conspicuous shift of $\phi(J)$ for the immersed case and $\phi(I)$ for the dry case toward lower values of $\phi$ with increasing surface roughness (or interparticle friction).

The semi-logarithmic plots of Fig.\,\ref{fig:MuPhi}(a) and (b) are particularly amenable to a close examination of the jamming transition, and in particular show that both $\phi$ and $\mu$ tend to finite measurable values, $\phi_c$ and $\mu_c$ respectively, at the jamming point. The critical (or maximum flowable) volume fraction, $\phi_c$, and friction coefficient, $\mu_c$, can be measured by fitting the data using a linear regression in $\sqrt{J}$ \cite{BoyeretalPRL2011} or in $I$ \cite{ForterrePouliquen2008} in the immersed or dry cases, respectively, as summarized in Table\,\ref{tab:coeff} and shown by the dashed lines of Fig.\,\ref{fig:MuPhi}(a) and (b). The critical volume fraction, $\phi_c$, happens to be similar within error bars in the immersed and dry cases and shows a decrease with increasing roughness (or interparticle friction). The value of $\phi_c$ measured for the SR spheres is found to be similar to those obtained ($\phi_c \approx 0.585$) in previous immersed experiments by \cite{BoyeretalPRL2011,Dagois-Bohyetal2015}  but differs with that obtained ($\phi_c \approx 0.625$) in the dry case by \cite{Falletal2015}. The values of $\mu_c$ for the immersed and dry cases are similar within error bars and do not significantly differ for the SR and HR spheres. They are slightly larger than those found previously ($\mu_c \approx 0.30-0.32$) in the immersed case by \cite{BoyeretalPRL2011,Dagois-Bohyetal2015} and much larger than that obtained ($\mu_c \approx 0.26-0.27$) in the dry case by \cite{Falletal2015}. The present value of 0.36 corresponds to a friction angle of $20$ degree which is close to the typical pile angles observed for spherical particles.

\begin{figure}
\begin{center}
\includegraphics[width=0.95\textwidth]{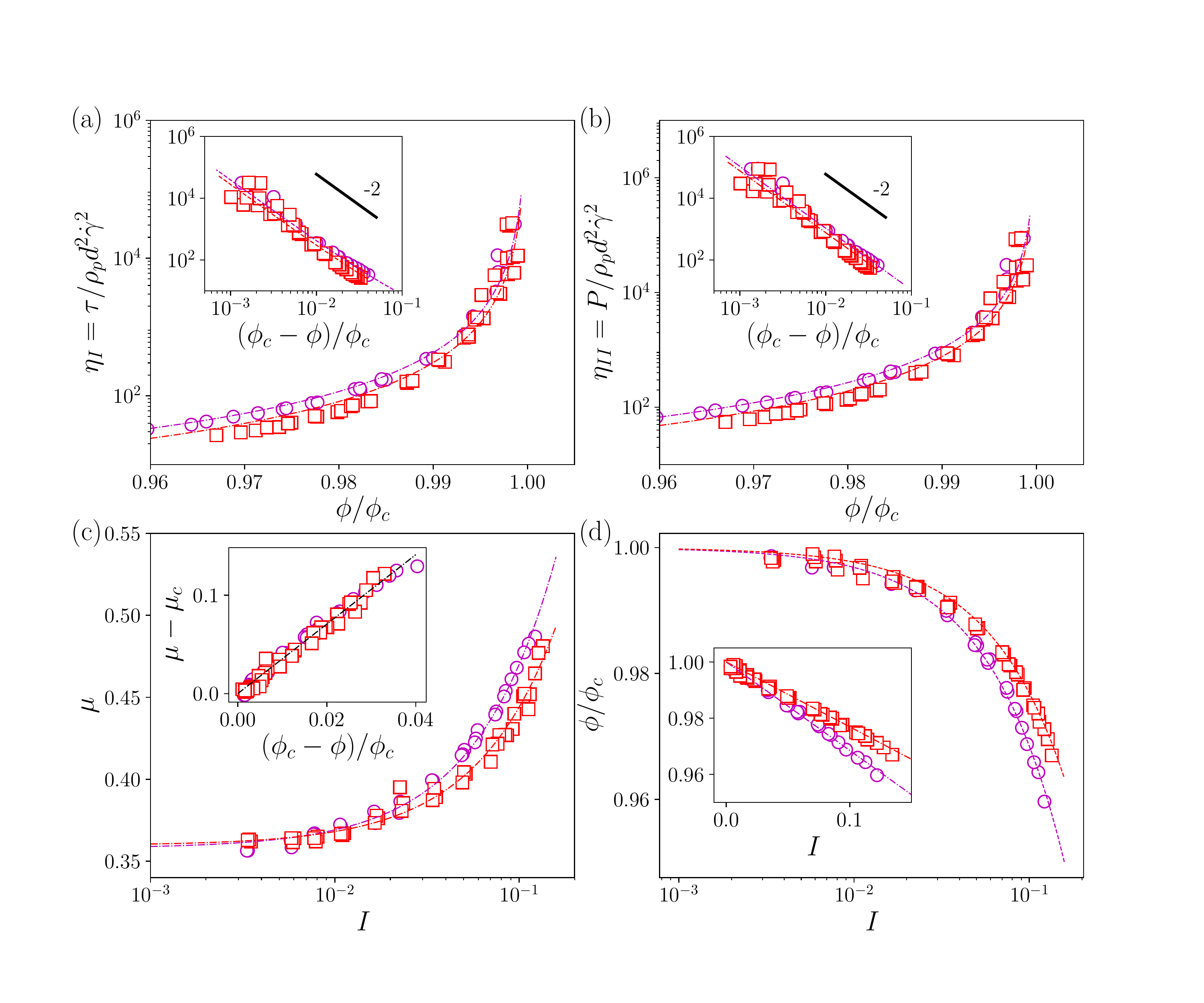}
\caption{Rheological data in the dry case: (a)~$\eta_I  = \tau/\rho_p d^2 \dot{\gamma}^2$ and (b)~$\eta_{II} =  P/\rho_p d^2 \dot{\gamma}^2$ versus $\phi/\phi_c$ as well as (c)~$\mu=\tau/P=\eta_s/\eta_n$ and (d)~$\phi$ versus $I$. The insets of graphs (a,b) and (d) are log-log and linear plots, respectively. The dashed lines correspond to the best fits using the models summarized in Table \ref{tab:coeff} and the linear variation $\mu(\phi) = \mu_c + a^i ( 1 - \phi/\phi_c)$ as a leading expansion. The inset of graph (c) shows the rescaled friction coefficient, $\mu-\mu_c$, versus the rescaled volume fraction, $1-\phi/\phi_c$, together with this leading expansion.}
\label{fig:dryRheo} 
\end{center}
\end{figure}

Fig.\,\ref{fig:MuPhi}(c) shows that $\mu$ presents a quasi-linear decrease in $\phi$ but with different slopes for the dry and immersed cases. By rescaling $\mu$ by $\mu-\mu_c$ and $\phi$ by $1-\phi/\phi_c$, good collapses of the data for the two different roughnesses are obtained in Fig.\,\ref{fig:MuPhi}(d) for the dry and immersed cases, respectively. A linear regression, $\mu(\phi) - \mu_c  \propto ( 1 - \phi/\phi_c)$, yields a decent fit of all the data with coefficients of proportionality $a^i \approx 3.52$ in the (dry) inertial case and $a^v \approx 8.05$ in the (immersed) viscous case, see the inset of Fig.\,\ref{fig:MuPhi}(d) and the inset of Fig. \ref{fig:dryRheo}(c). 

\begin{figure}
\begin{center}
\includegraphics[width=0.95\textwidth]{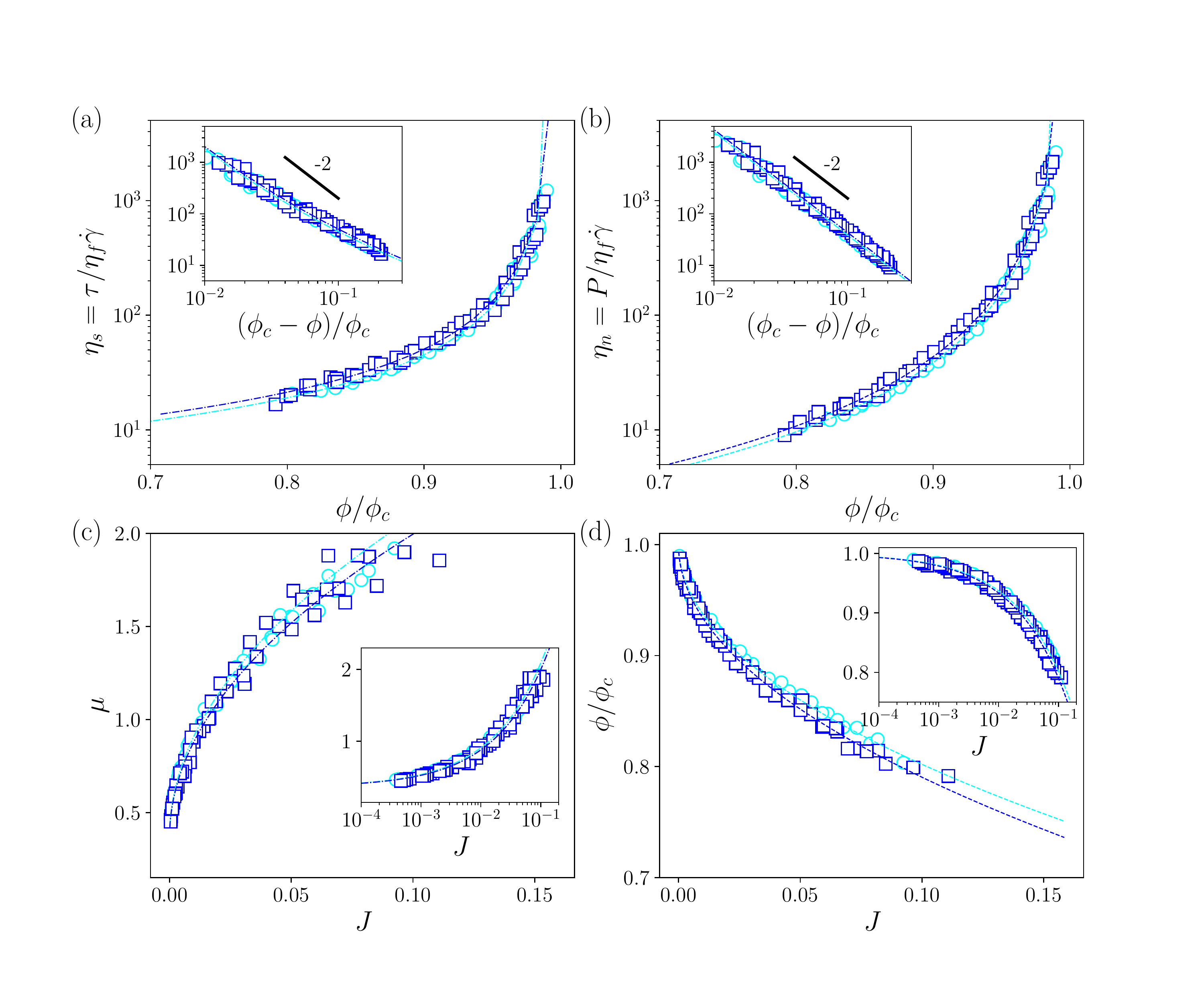}
\caption{Rheological data in the immersed case: (a)~$\eta_s  = \tau/\eta\dot{\gamma}$ and (b)~$\eta_n =  P/\eta\dot{\gamma}$ versus $\phi/\phi_c$ as well as (c)~$\mu=\tau/P=\eta_s/\eta_n$ and (d)~$\phi$ versus $J$. The insets of graphs (a,b) and (c,d) are log-log and semi-log plots, respectively. The dashed lines correspond to the best fits using the models summarized in Table \ref{tab:coeff} and the linear variation $\mu(\phi) = \mu_c + a^v ( 1 - \phi/\phi_c)$ as a leading expansion.}
\label{fig:wetRheo} 
\end{center}
\end{figure}

We now turn to a detailed examination of the rheological data obtained in the viscous (immersed) case in Fig. \ref{fig:wetRheo}. Rescaling $\phi$ by $\phi_c$ provides a complete collapse of all the data for both the SR and HR spheres. In addition to the frictional description displayed in Fig. \ref{fig:wetRheo}(c) and (d), we provide the classical viscosity description in Fig. \ref{fig:wetRheo}(a) and (b). The shear, $\eta_s  = \tau/\eta\dot\gamma$, and normal, $\eta_n =  P/\eta\dot\gamma$, relative viscosities increase with increasing $\phi$ and diverge asymptotically at the maximum flowable volume fraction, $\phi_c$. The log-log plot shown in the inset of Fig. \ref{fig:wetRheo}(b) reveals that $\eta_n$ diverges as $(1 - \phi/\phi_c)^{-2}$ near jamming. Note that since by construction $J=1/\eta_n$, the relation $\eta_n \propto (1 - \phi/\phi_c)^{-2}$ confirms the relation $(\phi_c - \phi) \propto J^{1/2}$ evidenced in Fig.\,\ref{fig:MuPhi}(b) for the immersed case. The shear viscosity, $\eta_s$, is determined by the rheological law $\eta_s(\phi)=\mu(\phi)\eta_n(\phi)$, where $\mu(\phi)$ is given by the leading expansion $\mu(\phi) = \mu_c + a^v ( 1 - \phi/\phi_c)$ introduced in Fig.\,\ref{fig:MuPhi}(d). The divergence in $(1 - \phi/\phi_c)^{-2}$ dominates close to jamming.

An analogous data representation for the inertial (dry) case is displayed in Fig.\,\ref{fig:dryRheo} and shows a decent collapse of the data when rescaling $\phi$ by $\phi_c$. While the collapse is excellent in the quasi-static regime, discrepancies arise as $I \gtrsim 10^{-2}$. The shear and normal stresses are normalized by the Bagnold scaling, $\rho_p d^2 \dot{\gamma}^2$ \cite{Bagnold1954}. This scaling defines two dimensionless functions of the volume fraction, $\eta_I= \tau/\rho_p d^2 \dot{\gamma}^2$ and $\eta_{II}=  P/\rho_p d^2 \dot{\gamma}^2$, which diverge when approaching the maximum volume fraction, $\phi_c$. The function $\eta_{II}$ is seen to diverge as $(1 - \phi/\phi_c)^{-2}$ near jamming as evidenced in the inset of Fig. \ref{fig:dryRheo}(b). This is consistent with the relation $(\phi_c - \phi) \propto I$ observed in Fig.\,\ref{fig:MuPhi}(b) for the dry case since $\eta_{II}=1/I^2$, see also the inset of Fig. \ref{fig:dryRheo}(d). The other dimensionless function is given by $\eta_I(\phi)=\mu(\phi)\eta_{II}(\phi)$ and has again a leading divergence in $(1 - \phi/\phi_c)^{-2}$ near jamming. A detailed display of $\mu(I)$, $\phi(I)$, and $\mu(\phi)$ is also provided in Fig. \ref{fig:dryRheo}(c) and (d). 

The rheological data of Figs.\,\ref{fig:MuPhi}, \ref{fig:wetRheo}, \ref{fig:dryRheo} are given as Supplemental Material.

\section{Discussion and Conclusions}
\label{sec:conclusion}

In this work, we have examined how particle roughness affects the rheological properties of particulate systems in both immersed and dry conditions. We have used two batches of spherical particles, a batch of ``slightly roughened'' regular spheres (SR) and another batch of ``highly roughened'' spheres (HR) produced by further roughening these regular spheres.
We have provided a detailed characterization of these particles, and more precisely determined their surface-roughness characteristics and their interparticle sliding and rolling friction coefficients using custom-made experimental devices. We have then performed pressure- and volume-imposed measurements of the rheology of theses two batches of spheres in the dense regime.

An important finding of the present study is the examination of the rheological behavior in the vicinity of the jamming transition in the dry and immersed cases. In the limit of vanishing shear rate, the critical values for the effective friction coefficient, $\mu_c$, and for the volume fraction, $\phi_c$, are found to be similar for suspensions and dry granular media. This seems to imply that hydrodynamic interactions are inconsequential and contact forces are prevailing in this quasi-static limit. A striking result is that $\mu_c$ is not significantly affected by particle roughness while $\phi_c$ decreases with increasing roughness. This later finding shows that the granular system needs to dilate more in order to flow when roughness is increased. Rescaling the volume fraction, $\phi$, by this maximum volume fraction, $\phi_c$, leads to an excellent collapse of all the data on master curves for the effective friction coefficient (or stress ratio), $\mu$, as well as for the shear and normal stresses, $\tau$ and $P$ respectively, normalized by a viscous scaling ($\eta\dot{\gamma}$) in the immersed case and by an inertial scaling ($\rho_p d^2 \dot{\gamma}^2$) in the dry case.

The excellent collapse of the experimental data by using the reduced volume fraction, $\phi/\phi_c$, has been previously observed for the shear viscosity in the suspension case \cite{GuazzelliPouliquen2018}, but without a controlled study of the influence of roughness and friction. Similar behavior has been also observed in numerical simulations when interparticle friction is increased \cite{Gallieretal2014,Marietal2014}. However these simulations match the present experimental $\phi_c$ for interparticle sliding friction coefficients ($\mu_{sf} \approx 0.5-1$) quite larger than those experimentally measured ($\mu_{sf}= 0.23$ for the SR spheres and $=0.37$ for the HR spheres). It is likely that rolling friction (with experimental coefficients $\mu_{sf}= 0.0043$ for the SR spheres and $=0.0070$ for the HR spheres) is somehow captured by using larger sliding friction coefficients as studied in granular media by \cite{Estradaetal2008}.

The present work not only yields the quasi-static values but also the asymptotic behaviors of $\mu$ and $\phi$ in the vicinity of the jamming transition. The departures of the effective friction coefficient and of the volume fraction from their static limits, $\mu-\mu_c$ and $\phi_c-\phi$, are found to vary as the square root of the viscous number in the immersed case, i.e. $\propto \sqrt{J}$, and linearly with the inertial number in the dry case, i.e.  $\propto I$. The present experimental data for the suspensions of the two batches of SR and HR spheres also reveal that both relative shear and normal viscosities, $\eta_s$ and $\eta_n$, present a similar algebraic divergence in $\approx (\phi_c-\phi)^{-2}$ as previously observed for a variety of diverse suspensions of frictional spheres \cite{BoyeretalPRL2011,GuazzelliPouliquen2018}. Interestingly, the analogous dimensionless functions obtained by normalizing the shear and normal stresses with an inertial scaling in the dry case, $\eta_I$ and $\eta_{II}$, present the same dominant algebraic divergence.

\begin{table}
\begin{ruledtabular}
\begin{tabular}{ccccccc}   
  &   relations & exponent    &  present &           predictions    & simulations  & simulations\\[-10pt]
  && & work & (frictionless) & (frictionless) & (frictional)\\
\hline
\multirow{2}{*}{\head{Viscous}}   &  $\mu- \mu_c \propto J^{\gamma_{\mu}}$  & $\gamma_{\mu}$ &  $0.5$    & $0.35$\cite{DeGiulietal2015}         & 0.32\cite{DeGiulietal2015}& $0.5$\cite{Trulssonetal2012,Amarsidetal2017}\\  
         \head{(immersed)}               & $\phi_c-\phi \propto J^{\gamma_{\phi}}$ &\multirow{2}{*}{$\gamma_{\phi}^{-1}$}  & \multirow{2}{*}{$2$}    & \multirow{2}{*}{$2.83$\cite{DeGiulietal2015}}        & \multirow{2}{*}{2.77\cite{DeGiulietal2015}} & \multirow{2}{*}{$2$\cite{Trulssonetal2012,Amarsidetal2017}}\\
            & $\eta_{s,n} \propto (\phi_c-\phi)^{-\gamma_{\phi}^{-1}}$ && &&1.5-1.6\cite{Marietal2014,Gallieretal2018} &2.3\cite{Marietal2014},1.9\cite{Gallieretal2018}\\
\hline                               
\multirow{2}{*}{\head{Inertial}}  &   $\mu - \mu_c \propto I^{\alpha_{\mu}}$ &  $\alpha_{\mu}$   & $1$  & $0.35$\cite{DeGiulietal2015}      & 0.38\cite{Peyneauetal2008} &1\cite{AzemaRadjai2014}\\  
                             \head{(dry)}                  &        $\phi_c-\phi \propto I^{\alpha_{\phi}}$& \multirow{2}{*}{$\alpha_{\phi}$}         &  \multirow{2}{*}{$1$}     & \multirow{2}{*}{$0.35$\cite{DeGiulietal2015}}          & \multirow{2}{*}{0.39\cite{Peyneauetal2008}} & \multirow{2}{*}{1\cite{AzemaRadjai2014}}   \\ 
     & $\eta_{I,II} \propto (\phi_c-\phi)^{-2\alpha_{\phi}^{-1}}$ & & & & &\\
\end{tabular}
\caption{Critical exponents found in the present work versus predicted and numerical values.}
\label{tab:exponents}
\end{ruledtabular}
\end{table}

The critical exponents found in the present experiments are compared in Table \ref{tab:exponents} to the theoretical predictions \cite{DeGiulietal2015} which have been obtained for frictionless spheres and to results of various numerical simulations performed in two dimensions \cite{Trulssonetal2012,DeGiulietal2015} as well as in three dimensions \cite{Marietal2014,Gallieretal2018,Peyneauetal2008,AzemaRadjai2014,Amarsidetal2017} for dry and immersed spheres with and without frictional interactions. They are in good agreement with the frictional simulations, in particular those performed in three dimensions. It is nonetheless difficult to assess what is the prevailing dissipation regime. The $J-\mu_{sf}$ diagram proposed by \cite{Trulssonetal2017} suggests that the immersed experiments lies between the rolling and frictional-sliding regimes while the $I-\mu_{sf}$ diagram proposed by  \cite{DeGiulietal2016} seems to indicate that the dry experiments belong to the frictional-sliding regime. However, the rolling friction is not accounted for in these studies. 

Finally, the present work may shed some light on the transition between the viscous and inertial regimes which is far from being fully understood and still a subject of debate \cite{Trulssonetal2012,DeGiulietal2015,Amarsidetal2017}. Assuming that this transition arises when the viscous and inertial stresses are matched, the crossover shear-rate is then $\dot{\gamma}_{v \rightarrow i} = (\eta_f/\rho_pd^2)\, \eta_s(\phi)/\eta_I(\phi)$. The similar divergence of the functions $\eta_s(\phi)$ and $\eta_I(\phi)$ found here seems to suggest that $\dot{\gamma}_{v \rightarrow i}$ is independent of the volume fraction close to the jamming transition. Further investigations are clearly necessary to understand thoroughly this viscous-inertial transition for dense frictional particles.

\begin{acknowledgments}
This work has been carried out thanks to the support of the ANR project `Dense Particulate Systems' (ANR-13-IS09-0005-01), the `Laboratoire d'Excellence M\'ecanique et Complexit\'e' (ANR-11-LABX-0092), the Excellence Initiative of Aix-Marseille University - A$^{\ast}$MIDEX (ANR-11-IDEX-0001-02) funded by the French Government `Investissements d'Avenir programme'. FT benefited from a fellowship of CONICYT (Engagement: 74170026).
\end{acknowledgments}

%==================================================================
%Bibliography here
\bibliography{References}

\end{document}